\input epsf
\documentclass[preprint,aps,pra,showpacs,floatfix]{revtex4}
\usepackage[english]{babel}
\usepackage{graphicx}
\usepackage{times}
\usepackage{nicefrac}
\usepackage{amsmath}
\usepackage{amsfonts}
\usepackage{amssymb}
\usepackage{amsthm}
\usepackage{epsf}
\usepackage{bm}
\usepackage{bbm}
\usepackage{times}
\usepackage{color}
\newcommand{\aZ}{\alpha Z}

\newcommand{\eps}{\varepsilon}
\newcommand{\balpha}{\mbox{\boldmath $\alpha$} }
\newcommand{\bmu}{\mbox{\boldmath $\mu$} }
\newcommand{\hfs}{hyperfine splitting }

\begin{document}
\title{
Radiative and interelectronic-interaction corrections
to the hyperfine splitting in highly charged B-like ions}
\author{N. S. Oreshkina${}^{1,2}$, D. A. Glazov${}^{1,2}$, A. V. Volotka${}^{2}$,
\\ V. M. Shabaev${}^{1}$, I. I. Tupitsyn${}^{1}$, and G. Plunien${}^{2}$}
\affiliation{
${}^1$ Department of Physics, St.Petersburg State University,
Oulianovskaya~1, Petrodvorets, St.Petersburg 198504, Russia \\
${}^2$ Institut f\"ur Theoretische Physik, TU Dresden,
Mommsenstra{\ss}e 13, D-01062 Dresden, Germany}
\begin{abstract} {The ground-state hyperfine splitting values of
high-$Z$ boronlike ions are calculated. Calculation of the
interelectronic-interaction contribution is based on a combination
of the $1/Z$ perturbation theory and the large-scale
configuration-interaction Dirac-Fock-Sturm method. The screened
QED corrections are evaluated utilizing an effective screening
potential approach. Total hyperfine splitting energies are presented
for several B-like ions of particular interest: ${}^{45}$Sc${}^{16+}$,
${}^{57}$Fe${}^{21+}$, ${}^{207}$Pb${}^{77+}$, and ${}^{209}$Bi${}^{78+}$.
For lead and bismuth the experimental values of the $1s$
hyperfine splitting are employed to improve significantly
the theoretical results by reducing the uncertainty due to
the nuclear effects.
}
\end{abstract}
\pacs{31.30.Jv, 31.30.Gs, 12.20.Ds}
\maketitle
%
\section{Introduction}

First accurate calculations \cite{int1,int2} of the hyperfine
splitting in highly charged ions were stimulated by astronomical
search in hot astrophysical plasma \cite{int4}. Later,
high-precision measurements of the ground-state hyperfine
splitting in heavy H-like ions were performed
\cite{ion1,ion2,ion3,ion4,bei-2001}. The main goal of such
experiments was to probe quantum electrodynamics (QED) in the strong
Coulomb field induced by a heavy nucleus. However, accurate
theoretical calculations (see Ref. \cite{shab-pr-2002} and
references therein) showed that the uncertainty of the theoretical
results, which mainly originates from the nuclear magnetization
distribution correction (the Bohr-Weisskopf effect), is comparable
with the QED correction. For this reason, any identification of
QED effects on the hyperfine splitting in heavy H-like ions turned out to be
unfeasible. It was shown, however, that this uncertainty can be
significantly reduced in a specific difference of the hyperfine
splitting values of H- and Li-like ions with the same nucleus
\cite{qed2001}. High-precision measurements of the hyperfine
splitting in heavy Li-like ions are presently in preparation
\cite{new_exp}.

The motivation for accurate calculations of the hyperfine splitting
in B-like ions is twofold. From one side, high-precision prediction
of the hyperfine splitting of B-like Fe may be important for astronomical
search \cite{sunyaev-priv}. From the other side, the study of the hyperfine
splitting in heavy B-like ions can be used to reduce the uncertainty
associated with the Bohr-Weisskopf effect in some specific difference
of the hyperfine splitting values for B- and Li-like ions or B- and H-like ions.
The origin of this reduction is essentially the same as for the
related $g$-factor values \cite{new-gf} and can be easily seen from
the approximate analytical expressions for the Bohr-Weisskopf
correction given in Refs. \cite{stat94,rev}.

In this article we calculate the ground-state hyperfine splitting of B-like ions.
The interelectronic-interaction correction of the first order in
$1/Z$ is evaluated within the rigorous QED approach. The
higher-order terms are calculated within the large-scale
configuration-interaction Dirac-Fock-Sturm method. QED corrections
are calculated using an effective potential approach in order to
account for the effect of screening.
The experimental values of the $1s$ hyperfine splitting
in H-like ${}^{207}$Pb${}^{81+}$ and ${}^{209}$Bi${}^{82+}$ are
employed to evaluate the Bohr-Weisskopf correction for
the corresponding B-like ions. Due to the correllation between
the values of this correction for $1s$ and $2p_{1/2}$ states
mentioned above, the uncertainties of the theoretical values
caused by the nuclear effects are strongly reduced.

Relativistic units ($\hbar=c=1$) and the Heaviside charge unit
($\alpha=e^2/(4\pi) $,  $e<0$) are used throughout the paper.

\section{Basic formulas and calculations}

Within the point-dipole approximation, the hyperfine interaction
is described by the Fermi-Breit operator
\begin{gather}
\label{Hmu}
  H_\mu=\frac{\mid e\mid}{4\pi}\frac{(\balpha\cdot
    [\bmu\times\mathbf{r}])}{r^3}
\,,
\end{gather}
where the vector $\balpha$ incorporates the Dirac matrices and
$\bmu$ is the nuclear magnetic moment operator. The ground-state
\hfs of a B-like ion can be written in the following form
\cite{rev,shab95}
\begin{equation}
\label{main}
\begin{split}
  \Delta E_\mu &= \frac{\alpha (\aZ)^3}{18}\frac{m}{m_p}
    \frac{\mu}{\mu_N}\frac{2I+1} {2I}mc^2
\\&
  \times \left[ A_\mu(\aZ)(1-\delta)(1-\varepsilon) +
    \frac{1}{Z}B_\mu(\aZ)+\frac{1}{Z^2}C_\mu(Z,\aZ)+x_{\rm rad} \right]
\,,
\end{split}
\end{equation}
where $m$ is the electron mass, $m_p$ is the proton mass, $\mu_N$
is the nuclear magneton, and $I$ is the nuclear spin.
$A_\mu(\alpha Z)$ is the one-electron relativistic factor
\begin{equation}
\label{ab}
  A_\mu(\alpha Z) = \frac{6[2(1+\gamma)-\sqrt{2(1+\gamma)}\,]}
    {(1+\gamma)^2\gamma(4\gamma^2-1)}
\,,
\end{equation}
where $\gamma = \sqrt{1-(\alpha Z)^2}$, $\delta$ is the nuclear-charge-distribution
correction, and $\varepsilon$ is the one-electron nuclear-magnetization-distribution
correction (the Bohr-Weisskopf effect). The terms $B_\mu(\aZ)/Z$ and $C_\mu(Z,\aZ)/Z^2$
determine the interelectronic-interaction correction to the first and higher orders
in $1/Z$, respectively. The $x_{\rm rad}$ term stands for the QED correction.

Finite-nuclear-size correction $\delta$ can be calculated both
analytically \cite{stat94,volotka03} and numerically. In the present work,
it is obtained numerically by solving the Dirac equation with the
Fermi model for the nuclear charge distribution. The Bohr-Weisskopf
correction $\varepsilon$ is calculated within the single-particle
model as described in Refs. \cite{int1,int2,shab97}. Apart from
this direct evaluation, at the end of this section we also
derive the $\varepsilon$ values for ${}^{207}$Pb${}^{77+}$
and ${}^{209}$Bi${}^{78+}$ using the experimental results
for the $1s$ hyperfine splitting. The nuclear root-mean-square radii
are taken from Ref. \cite{angeli} and the nuclear magnetic moments
are taken from Ref. \cite{nuc-par}.

%
\begin{figure}
\includegraphics{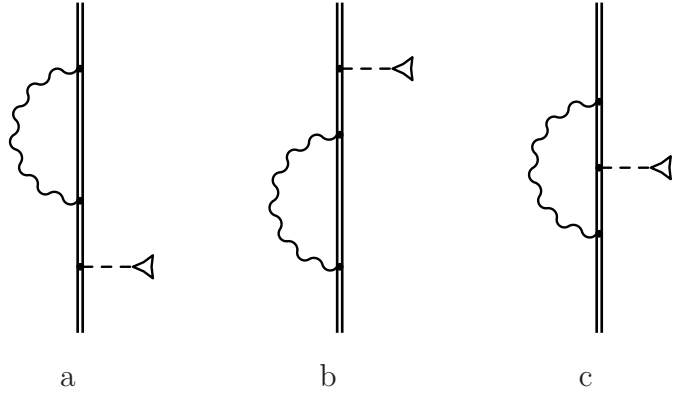} \caption {Feynman diagrams
representing the self-energy correction to the hyperfine
splitting. The double line indicates the bound-electron propagator
and the dashed line terminated with the triangle denotes the
hyperfine interaction.}\label{SE}
\end{figure}
\begin{figure}
\includegraphics{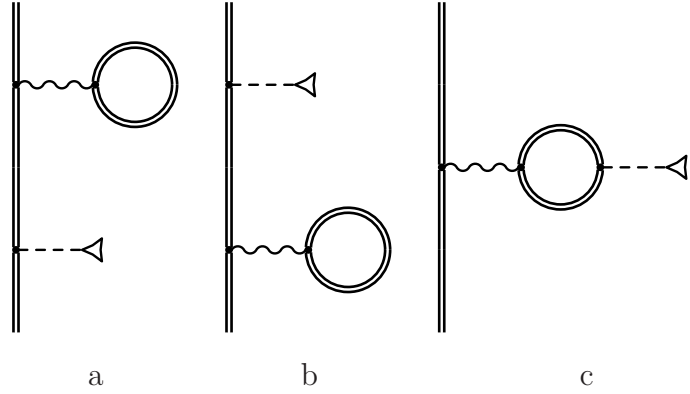} \caption {Feynman diagrams
representing the vacuum-polarization correction to the hyperfine
splitting. Notations are the same as in Fig.~\ref{SE}.}\label{VP}
\end{figure}
The QED correction to the hyperfine splitting of the first order in $\alpha$
consists of two parts, self-energy correction and vacuum-polarization correction.
The self-energy correction (Fig.~\ref{SE}) is the sum of irreducible, reducible,
and vertex parts,
\begin{equation}
  \Delta E_{\rm SE} = \sum_{M_Im}\sum_{M'_Im'} C^{FM_F}_{IM'_Ijm'}
    C^{FM_F}_{IM_Ijm}\chi^{+}_{IM'_I} (M_{\rm irr} + M_{\rm red} +
    M_{\rm ver})\chi_{IM_I}
\,,
\end{equation}
where $C^{FM_F}_{IM_Ijm}$ are the Clebsch-Gordan coefficients and
$\chi_{IM_I}$ is the nuclear wave function. The irreducible part
is given by the expression
\begin{equation}
\label{irr}
  M_{\rm irr} = \sum_{n}^{\eps_n\neq\eps_a} \frac {\langle a'
    |(\Sigma(\eps_a)-\gamma^0\delta m)|n\rangle \langle n|H_\mu|a
    \rangle} {\eps_a-\eps_n}
\,.
\end{equation}
Here $|a\rangle$ and $|a'\rangle$ are the states of the valence
electron with angular momentum projections $m$ and $m'$,
respectively, $\eps_n$ is the energy of the state $|n\rangle$ in
the binding potential under consideration, $H_\mu$ is the
magnetic-dipole hyperfine-interaction operator (\ref{Hmu}), and
$\delta m$ is the mass counterterm. $\Sigma(\eps)$ denotes the
unrenormalized self-energy operator defined as
\begin{equation}
  \langle a |\Sigma(\eps)|b\rangle =
    \frac{i}{2\pi}\int_{-\infty}^{\infty}d\omega \sum_n \frac {\langle
    an|I(\omega)|nb \rangle}{\eps - \omega - \eps_n(1-i0)}
\,,
\end{equation}
where $I(\omega,x_1,x_2) = e^2\alpha^\mu\alpha^\nu D_{\mu\nu}(\omega, x_1-x_2)$
with the Dirac matrices $\alpha^\mu = (1,\balpha)$, and the photon propagator $D_{\mu\nu}$.
To separate the ultraviolet divergencies, the expression (\ref{irr})
is decomposed into zero-, one-, and many-potential terms. The zero-potential
$M^{(0)}_{\rm irr}$ and one-potential $M^{(1)}_{\rm irr}$ terms are calculated
in momentum space using the formulas from Ref. \cite{snyd}. The residual part
of $M_{\rm irr}$, the so-called many-potential term $M^{(2+)}_{\rm irr}$,
is calculated in coordinate space. The expressions for the reducible and
the vertex parts read
\begin{align}
  M_{\rm red} &= \langle a'|H_\mu|a \rangle\langle a
    |\frac{d}{d\eps} \Sigma(\eps) \mid_{\eps = \eps_a} |a\rangle
\,,
\\
  M_{\rm ver} &= \frac{i}{2\pi} \int_{-\infty}^{\infty} d\omega
    \sum_{n_1,n_2} \frac {\langle a'n_2 |I(\omega)|n_1a\rangle \langle
    n_1|H_\mu|n_2 \rangle} {(\eps_a-\omega-\eps_{n_1}(1-i0))
    (\eps_a-\omega-\eps_{n_2}(1-i0))}
\,.
\end{align}
Both reducible and vertex parts are ultraviolet-divergent, whereas
the sum $M_{\rm vr} = M_{\rm red} + M_{\rm ver}$ is finite.
Following Refs. \cite{persson, blundell:97,se2,sap-2001}, we
separate out zero-potential term $M_{\rm vr}^{(0)}$ and evaluate
it in momentum space. The remaining many-potential term $M_{\rm
vr}^{(1+)}$ is calculated in coordinate space as a point-by-point
difference between the contributions with bound and free
propagators in the self-energy loop.

The angular integration and the summation over intermediate angular
projections in the many-potential terms $M_{\rm irr}^{(2+)}$ and
$M_{\rm vr}^{(1+)}$ are carried out in a standard manner.
The many-potential terms involve infinite summation over relativistic
angular quantum number $\kappa=\pm(j+1/2)$. The summation is terminated
at a maximum value $|\kappa|=10$, while the residual part
of the sum is evaluated by the least-square inverse-polynomial fitting.
For any given $\kappa$ the summation over the Dirac spectrum is
performed utilizing the dual-kinetic-balance (DKB) approach \cite{dkb}
involving basis functions constructed from B-splines.
The finite distributions of the nuclear charge and the nuclear magnetic
moment are taken into account.

%
The electric-loop part of the vacuum-polarization correction
(parts (a) and (b) of Fig.~\ref{VP}) is evaluated within an
effective screening potential employing the DKB method. Explicit formulas
for the Uehling potential can be found, e.g., in Refs. \cite{shab-pr-2002,mohr:98:PREP}.
For the evaluation of the electric-loop Wichmann-Kroll potential we
consult the approximate formulas derived in Ref. \cite{fainshtein}.
The magnetic-loop part (diagram in Fig.~\ref{VP} (c)) in the Uehling
approximation was considered in Ref. \cite{schneider:94:PRA}. We evaluate
this term in the presence of an effective screening potential with
account for the finite distribution of the nuclear charge and the nuclear
magnetic moment. The value of the magnetic-loop Wichmann-Kroll part
is estimated to be rather small. It was obtained by analyzing its
relative contribution for $s$ states \cite{sun-1998,artemyev:2001}.
However, in Table~\ref{all} we include its estimation to the total QED
term for the case of ${}^{207}$Pb${}^{77+}$ and ${}^{209}$Bi${}^{78+}$
ions.

%
%

In the present work, the QED corrections are evaluated based on four
different spherically symmetric binding potentials $V_{\rm eff}(r)$
which account for the interelectronic interaction between
the valence $2p_{1/2}$ electron and the core electrons of the
closed $(1s)^2(2s)^2$ shells. The simplest choice of $V_{\rm eff}$
is the core-Hartree (CH) potential
\begin{equation} \label{ch}
  V_{\rm eff}(r) = V_{\rm nuc}(r) + \alpha \int_{0}^{\infty}dr'
    \frac{1}{r_>}\rho_c(r')
\,,
\end{equation}
where $V_{\rm nuc}$ is the nuclear potential and $\rho_c$ is the
density of the core electrons. The screening potential derived
from the density-functional theory reads
\begin{equation} \label{dh}
  V_{\rm eff}(r) = V_{\rm nuc}(r) + \alpha \int_{0}^{\infty}dr'
    \frac{1}{r_>}\rho_t(r') - x_a \frac{\alpha}{r} \biggl(
    \frac{81}{32\pi^2}r\rho_t(r) \biggr)^{1/3}
\,.
\end{equation}
Here $\rho_t$ is the total electron density, including $(1s)^2(2s)^2$
shells and $2p_{1/2}$ electron. The parameter $x_a$ is varied from 0 to 1.
The cases of $x_a=0,2/3$, and $1$ correspond to the Dirac-Hartree (DH),
the Kohn-Sham (KS), and the Dirac-Slater (DS) potentials, respectively.
At large $r$ expression (\ref{dh}) should be replaced by
\begin{equation}
  V_{\rm eff}(r) = -\frac{\alpha(Z-4)}{r}
\end{equation}
to provide the proper asymptotic behavior. The self-consistent potential
is generated by iterations.

The results obtained for the QED correction for the pure nuclear
potential and the screening potentials are presented in Table \ref{rad}.
The first column corresponds to the values of the QED correction for
the nuclear potential $V_{\rm nuc}$. The second column gives the values
of the screened QED correction for the core-Hartree potential \eqref{ch},
the other columns present the values of the screened QED correction for
the Dirac-Hartree potential (eq. \eqref{dh}, $x_a = 0$), the Kohn-Sham
potential (eq. \eqref{dh}, $x_a = 2/3$), and the Dirac-Slater potential
(eq. \eqref{dh}, $x_a = 1$), respectively. The results for the pure
nuclear potential agree with those presented in Ref. \cite{sap}.
\begin{table}
\caption{The QED correction $x_{\rm rad}$ for the pure nuclear
potential $V_{\rm nuc}$ and for the effective screening potentials
(core-Hartree, Dirac-Hartree, Kohn-Sham and Dirac-Slater
potentials, respectively).}
\label{rad}
\tabcolsep10pt
\begin{tabular}{|l|r@{}l|r@{}l|r@{}l|r@{}l|r@{}l|}
\hline
$Z$   &
\multicolumn{2}{c|}{$V_{\rm nuc}$} &
\multicolumn{2}{c|}{CH}            &
\multicolumn{2}{c|}{DH}            &
\multicolumn{2}{c|}{KS}            &
\multicolumn{2}{c|}{DS}            \\
\hline
15    &   0.&00047 &   0.&00030 &   0.&00027 &   0.&00031 &   0.&00033 \\
21    &   0.&00041 &   0.&00032 &   0.&00029 &   0.&00032 &   0.&00033 \\
26    &   0.&00035 &   0.&00030 &   0.&00028 &   0.&00030 &   0.&00032 \\
37    &   0.&00020 &   0.&00020 &   0.&00020 &   0.&00020 &   0.&00020 \\
49    &$-$0.&00005 &   0.&00001 &   0.&00002 &   0.&00001 &$-$0.&00000 \\
57    &$-$0.&00029 &$-$0.&00018 &$-$0.&00016 &$-$0.&00019 &$-$0.&00021 \\
67    &$-$0.&00073 &$-$0.&00056 &$-$0.&00053 &$-$0.&00058 &$-$0.&00060 \\
75    &$-$0.&0013  &$-$0.&0010  &$-$0.&0010  &$-$0.&0011  &$-$0.&0011  \\
82    &$-$0.&0020  &$-$0.&0017  &$-$0.&0017  &$-$0.&0017  &$-$0.&0019  \\
83    &$-$0.&0023  &$-$0.&0018  &$-$0.&0018  &$-$0.&0019  &$-$0.&0020  \\
\hline
\end{tabular}
\end{table}

In the calculations performed the effect of screening on QED
correction is taken into account only in the local effective
potential approximation. We estimate that for the $p_{1/2}$
valence state the uncertainty due to this approximation amounts to
about $50\%$ for $Z=15$ and decreases rapidly as $Z$ increases.
Evaluation of the screened QED correction within the rigorous QED
approach is presently underway.

The interelectronic-interaction correction of the first order in
$1/Z$ defined by the function $B_\mu(\aZ)$ can be calculated
within the rigorous QED approach. Such calculations for the ground
and first excited states of Li-like ions were performed in Refs.
\cite{shab95,mshab-1999,kor-ore}. The formulas derived there can
easily be adopted for B-like ions by regarding the closed
(1s)${}^2$ and (2s)${}^2$ shells as belonging to the vacuum state
\cite{shab-pr-2002}. The results of the numerical evaluation by
these formulas are presented in Table \ref{iei}. The function
$B_{\mu}^{(0)}(\aZ)$ indicates the values obtained for the
point-nucleus case using the method of the generalized virial
relations for the Dirac equation \cite{virial}. The function
$B_{\mu}^{\rm (NS)}(\aZ)$ takes into account the nuclear charge
distribution effect. Its numerical calculation is performed using
the DKB approach. The function $B_{\mu}^{\rm (BW)}(\aZ)$
incorporates also the Bohr-Weisskopf effect.
\begin{table}
\caption{First-order interelectronic-interaction correction for
the ground-state of B-like ions. $B_{\mu}^{(0)}(\aZ)$ and
$B_{\mu}^{\rm (NS)}(\aZ)$ correspond to the point and extended
charge nucleus, respectively. $B_{\mu}^{\rm (BW)}(\aZ)$ includes
the BW correction.}\vspace{0.5cm}
\label{iei}
\tabcolsep10pt
\begin{tabular}{|l|r@{}l|r@{}l|r@{}l|}
\hline
$Z$  &
\multicolumn{2}{c|}{$B_{\mu}^{(0)}(\aZ)$} &
\multicolumn{2}{c|}{$B_{\mu}^{\rm (NS)}(\aZ)$} &
\multicolumn{2}{c|}{$B_{\mu}^{\rm (BW)}(\aZ)$} \\
\hline
15   &$-$5.&90979  &$-$5.&9097  &$-$5.&9097  \\
21   &$-$6.&11189  &$-$6.&1117  &$-$6.&1116  \\
26   &$-$6.&34230  &$-$6.&3417  &$-$6.&3413  \\
37   &$-$7.&09270  &$-$7.&090   &$-$7.&090   \\
49   &$-$8.&44727  &$-$8.&434   &$-$8.&431   \\
57   &$-$9.&83459  &$-$9.&801   &$-$9.&798   \\
67   &$-$12.&4859  &$-$12.&37   &$-$12.&35   \\
75   &$-$15.&8639  &$-$15.&56   &$-$15.&51   \\
82   &$-$20.&4837  &$-$19.&75   &$-$19.&51   \\
83   &$-$21.&3367  &$-$20.&49   &$-$20.&42   \\
\hline
\end{tabular}
\end{table}

The higher-order term $C_\mu(Z,\aZ)/Z^2$ is evaluated using the
large-scale configuration-interaction Dirac-Fock-Sturm (CI-DFS)
method \cite{tup-dfs1,tup-dfs2,dfs}. The many-electron wave
function $\Psi(\gamma J)$ with the total angular momentum $J$ and
other quantum numbers $\gamma$ is expanded in terms of a large
number of the configuration state functions (CSFs) with the same $J$,
\begin{equation}\label{dfs}
  \Psi(\gamma J) = \sum_{\alpha}c_{\alpha}\Phi_{\alpha}(J)
\,.
\end{equation}
For each relativistic atomic configuration the CSFs $\Phi_{\alpha}(J)$
are eigenfunctions of $J^2$ and can be obtained as linear
combinations of the Slater determinants corresponding to this configuration.
The set of the CSFs in the expansion (\ref{dfs}) was generated including all
single, double and triple excitations. The Slater determinants are constructed
from one-electron four-component Dirac spinors (orbitals). For the occupied
shells these orbitals were obtained by the multiconfiguration Dirac-Fock method.

The CI-DFS method allows us to calculate the interelectronic-interaction
correction to all orders in $1/Z$ within the Breit approximation, whereas
the $B_\mu(\aZ)/Z$ term is obtained within the rigorous QED approach
as described above. In order to combine these approaches, we subtract
the value of the $1/Z$ term calculated within the Breit approximation
from the CI-DFS result. In this way we obtain the $C_\mu(Z,\aZ)/Z^2$ contribution.


Table \ref{all} presents the individual contributions and the
total theoretical results for the ground-state \hfs in B-like ions
of particular interest. It can be seen that for heavy ions
the uncertainties of the total theoretical values are completely
determined by the Bohr-Weisskopf effect. These uncertainties can be
strongly reduced employing the experimental values for the hyperfine
splitting in the corresponding H-like ions. Following the works
\cite{qed2001,rev} we use the experimental values for the ground state
hyperfine splitting, $\Delta E^{(1s)}_{\rm exp} = 1.2159(2)$ eV
for H-like ${}^{207}$Pb${}^{81+}$ \cite{ion4}
and $\Delta E^{(1s)}_{\rm exp} = 5.0840(8)$ eV
for H-like ${}^{209}$Bi${}^{82+}$ \cite{ion1},
to extract the Bohr-Weisskopf corrections for the $1s$ state
employing the theoretical values for all other contributions
from Ref. \cite{shab-pr-2002}. Considering different models
for the nuclear magnetization distribution, we have found that
the ratio of the Bohr-Weisskopf corrections is rather stable,
$\varepsilon^{(2p)}_{\rm Pb}/\varepsilon^{(1s)}_{\rm Pb}=0.287(2)$ for Pb and
$\varepsilon^{(2p)}_{\rm Bi}/\varepsilon^{(1s)}_{\rm Bi}=0.295(2)$ for Bi.
It allows us to deduce the following values for B-like ions:
$\varepsilon^{(2p)}_{\rm Pb}=0.0119(2)$ and $\varepsilon^{(2p)}_{\rm Bi}=0.00437(15)$.
The corresponding contributions to the hyperfine splittings
are $-0.83(1)$ meV and $-1.25(4)$ meV, respectively.
The total theoretical values, which include also the modification
of the Bohr-Weisskopf effect to the interelectronic-interaction correction,
amount to $62.24(2)$ meV for ${}^{207}$Pb${}^{77+}$
and $257.84(5)$ meV for ${}^{209}$Bi${}^{78+}$.
It should be stressed that the uncertainty of these values is not equal
to the sum of the uncertainties of the individual contributions. This
is due to the fact, that the total hyperfine splitting value found
in this way is sufficiently stable with respect to possible variations
of the nuclear charge radius and the magnetic moment \cite{rev}.
\begin{table}
\caption{Individual contributions to the ground-state \hfs of B-like ions, in meV.
The total error bars indicated do not include the nuclear magnetic moment
uncertainties \cite{nuc-par}.}
\vspace{0.5cm}
\label{all}
\begin{tabular}
{|l|r@{}l|r@{}l|r@{}l|r@{}l|}
\hline
&
\multicolumn{2}{c|}{${}^{45}$Sc${}^{16+}$}  &
\multicolumn{2}{c|}{${}^{57}$Fe${}^{21+}$}  &
\multicolumn{2}{c|}{${}^{207}$Pb${}^{77+}$} &
\multicolumn{2}{c|}{${}^{209}$Bi${}^{78+}$}
\\
Effect       &
\multicolumn{2}{c|}{$\frac{\mu}{\mu_N}=4.7565$}    &
\multicolumn{2}{c|}{$\frac{\mu}{\mu_N}=0.090623$}  &
\multicolumn{2}{c|}{$\frac{\mu}{\mu_N}=0.59258$}   &
\multicolumn{2}{c|}{$\frac{\mu}{\mu_N}=4.1106$}
\\
\hline
Dirac value         &   2.&3126    &   0.&15010    &  71.&89      &  296.&35     \\
Finite nuclear size &$-$0.&0001    &$-$0.&00001    &$-$2.&18(1)   & $-$9.&84(5)  \\
Bohr-Weisskopf      & & & & & & & & \\
(direct calculation)&   0.&0000    &$-$0.&00001    &$-$0.&84(8)   & $-$0.&97(34) \\
Interelectronic interaction, $1/Z$
                    &$-$0.&6424    &$-$0.&03406    &$-$6.&82      & $-$28.&17    \\
Interelectronic interaction, & & & & & & & &\\
$1/Z^2$ and higher orders
                    &   0.&0408(9) &   0.&00182(3) &   0.&24(1)   &    0.&98(3)  \\
QED                 &   0.&0007    &   0.&00004    &$-$0.&06(1)   & $-$0.&26(3)  \\
Total               &   1.&7116(9) &   0.&11788(3) &  62.&23(8)   &  258.&09(35) \\
\hline
Bohr-Weisskopf      & & & & & & & & \\
(from the $1s$ experiment)
                    &     &        &     &         &$-$0.&83(1)   & $-$1.&25(4)  \\
Total               &     &        &     &         &  62.&24(2)   &  257.&84(5)  \\
\hline
\end{tabular}
\end{table}
%
\section{Conclusion}
%
In this paper we have calculated the ground-state hyperfine
splitting of high-$Z$ boronlike ions.
The interelectronic-interaction correction is evaluated utilizing
the $1/Z$ perturbation theory and the large-scale configuration-interaction
Dirac-Fock-Sturm method. The radiative corrections are calculated
in the presence of an effective potential that partly accounts for the screening effect.
It is shown that the Bohr-Weisskopf effect for heavy boronlike ions
can be calculated with a high precision utilizing the experimental value
for the ground-state hyperfine splitting in the corresponding hydrogenlike
ions. As a result, the most accurate theoretical predictions for the
hyperfine splitting values of B-like Sc, Fe, Pb, and Bi are obtained.
%
\section*{Acknowledgments}
%
The work was supported by RFBR (Grant No. 07-02-00126a) and by
INTAS-GSI (Grant No. 06-1000012-8881). The work of N.S.O. and
D.A.G. was supported by DAAD. N.S.O. also acknowledges the support by
the ``Dynasty'' foundation. D.A.G., A.V.V., I.I.T., and G.P.
acknowledge financial support from the BMBF, DFG and GSI.
\end{document}